\documentclass[prd,showpacs,twocolumn,preprintnumbers,amsmath,amssymb,superscriptaddress,floatfix,nofootinbib]{revtex4-2}%

\usepackage{graphicx}
\usepackage{bm}
\usepackage{amsmath}
\usepackage{amsfonts}
\usepackage{amssymb}
\usepackage{color}
\usepackage{multirow}
\usepackage{graphicx}
\usepackage{subfigure}
\usepackage[colorlinks, citecolor=blue,anchorcolor=red,menucolor=red, linkcolor=red,filecolor=red,runcolor=red,urlcolor=blue,frenchlinks=red]{hyperref}

\newcommand{\xiu}[1]{\textcolor{black}{#1}}

\begin{document}
\title{Role of the $\Sigma(1430)(1/2^-)$ in the $J/\psi \to \Lambda \bar{\Lambda} \pi^0$ reaction}
	
\author{Yu-Shan Ren}\email{ryskxky@yeah.net}
\affiliation{School of Physics, University of Electronic Science and Technology of China, \\Chengdu 610054, China}
\affiliation{Departamento de Física Teórica and IFIC, Centro Mixto Universidad de Valencia-CSIC Institutos de Investigación de Paterna, 46071 Valencia, Spain}
\vspace{0.5cm}

\author{Wen-Tao Lyu}\email{lvwentao9712@163.com}
\affiliation{School of Physics, Zhengzhou University, Zhengzhou 450001, China}
\affiliation{Departamento de Física Teórica and IFIC, Centro Mixto Universidad de Valencia-CSIC Institutos de Investigación de Paterna, 46071 Valencia, Spain}
\vspace{0.5cm}
		
\author{Eulogio Oset}\email{oset@ific.uv.es}
\affiliation{Departamento de Física Teórica and IFIC, Centro Mixto Universidad de Valencia-CSIC Institutos de Investigación de Paterna, 46071 Valencia, Spain}
\vspace{0.5cm}

\begin{abstract}
We study the $J/\psi \to \bar{\Lambda} \Lambda \pi^0$ reaction, an \xiu{isospin-violating} reaction, by looking at the trios of \xiu{a} baryon-antibaryon-pseudoscalar meson that conform to a singlet of $\text{SU}(3)$ in the $u, d, s$ quarks. These terms conserve isospin; however, once the final-state interactions of meson-baryon and meson-antibaryon are taken into account, \xiu{the} isospin is violated due to the different masses of particles within the same isospin multiplets. Since the reaction is tied to the interaction of particles, only resonances that are dynamically generated by these interactions show up in the reaction. In this sense, our approach produces the $\Sigma(1430)(1/2^-)$ state \xiu{ but not the $\Sigma(1385)(3/2^+)$ state}. Comparing with the BESIII data we observe that, within the limited statistics of the experiment, the data show a structure {around $M_{\pi\Lambda} = 1430$ MeV}  that is reproduced by the theory, and no signal is seen for the $\Sigma(1385)(3/2^+)$ as the theory predicts. We call for a future update of the experiment once better statistics become available.

\end{abstract}
	
	\date{\today}
	
	\maketitle
	
\section{Introduction}\label{sec1}
In the Review of Particle Physics (RPP)~\cite{ParticleDataGroup:2024cfk}, one has the ground state $\Sigma(1/2^+)$ with a mass $M_{\Sigma^+}=1189.37\text{ MeV}$, followed by the $\Sigma(1385)(3/2^+)$ and the next state, the $\Sigma(1580)(3/2^-)$, followed by the $\Sigma(1620)(1/2^-)$.
In the relativized quark model with \xiu{the} chromodynamics of Capstick and Isgur~\cite{Capstick:1986ter}, the two ground states $\Sigma(1189)(1/2^+)$ and $\Sigma^*(1385)(3/2^+)$ are reasonably obtained with masses of $1190\text{ MeV}$ and $1370\text{ MeV}$, respectively.
For the $P$-wave states, the authors obtain $M_{\Sigma(1/2^-)}=1630\text{ MeV}$ and $M_{\Sigma(3/2^-)}=1655\text{ MeV}$. 

Interestingly, neither the RPP nor the quark model of Ref.~\cite{Capstick:1986ter} reports a $1/2^-$ $\Sigma$ state with energy below 1620 MeV.
The advent of chiral Lagrangians for baryons~\cite{Ecker:1994gg} and the subsequent developments on chiral unitary theory~\cite{Oset:1997it, Lutz:2001yb, Oller:2000fj, Jido:2003cb, Hyodo:2011ur, Jido:2001nt, Garcia-Recio:2002yxy} \xiu{shed new light on the subject}, showing that there are baryon states which are not a consequence of the interaction of three quarks, but stem from the interaction of mesons with baryons to form baryonic molecules.
One of the emblematic examples is the appearance of two $\Lambda(1405)(1/2^-)$ states~\cite{Oller:2000fj, Jido:2003cb}, one around 1380 MeV and the other one around 1420 MeV, now officially  {included} in the RPP. 
Concerning the $\Sigma$ states, the work of Ref.~\cite{Jido:2003cb} showed also another surprise. In a SU(3) symmetric world, the interaction of pseudoscalar mesons with baryons of the SU(3) octet gave rise to a singlet state and an octet state. 
When the SU(3) symmetry was gradually broken to reach a realistic situation where the masses of the particles of the same SU(3) multiplet had their physical values, the octet split into four branches, two with isospin $I=0$, and two with $I=1$. 
One of the $I=0$ octet branches and the singlet branch gave rise to the two $\Lambda(1405)$ states{,} and two $1/2^-$ $I=1$ states emerged, one around 1600 MeV that could be associated to the $\Sigma(1620)(1/2^-)$ \xiu{and} another one that appeared at the threshold of $\bar{K}N$, around 1430 MeV. 
This latter state was slightly bound or slightly unbound {depending on the input used}, giving rise to a cusp structure.

Information from CLAS photoproduction data with the $\gamma p \to K \pi \Sigma$ reaction~\cite{CLAS:2013rjt} was analyzed within the framework of the chiral unitary approach~\cite{Roca:2013cca} and the parameters obtained from a fit to the data led to a slightly unbound $\Sigma(1/2^-)$ state with a cusp-like amplitude peaking at the $\bar{K}N$ threshold. 
There have been many studies and proposals to search for this state experimentally (see a detailed list in Ref.~\cite{Lin:2025pyk}), but the experimental finding was done by the Belle collaboration~\cite{Belle:2022ywa} where a clean peak was observed in the $\pi^+\Lambda$ and $\pi^-\Lambda$ mass distributions of the $\Lambda_c^+ \to \Lambda \pi^+ \pi^+ \pi^-$ reaction, which was suggested in Ref.~\cite{Xie:2018gbi}.

In the present work, we study the $J/\psi \to \Lambda \bar{\Lambda} \pi^0$ reaction, which violates isospin. Studies of isospin violation~\cite{Achasov:1979xc,
 Hanhart:2007bd} show that the most important mechanisms for this violation come from rescattering effects due to the \xiu{noncancellation} of loops because of different masses of particles belonging to the same isospin multiplets. We adopt the same philosophy here.
Our starting point is to take the $J/\psi$ state as a singlet of SU(3). Then, we take all the structures with a baryon, an antibaryon\xiu{,} and a pseudoscalar meson\xiu{,} which form {a} SU(3) singlet, and we choose from there all terms that can lead to the final $\Lambda \bar{\Lambda} \pi^0$ through rescattering. 
Certainly, the $\Lambda \bar{\Lambda} \pi^0$ is not one of the original terms because the SU(3) singlet only contains singlets of isospin, but the $\Lambda \bar{\Lambda} \pi^0$ will finally be formed from rescattering mechanisms and \xiu{the} sum of terms that would cancel in the limit of equal masses in the same isospin multiplets \xiu{but do not cancel} when the physical masses are considered.

The other interesting thing about the reaction is that if resonances appear in the invariant mass distributions, they must correspond to states that appear because of the interaction of hadrons, what we call dynamically generated resonances by definition. 
In that sense, we expect to find a signal of the $\Sigma(1430)(1/2^-)$, which for us is generated from the interaction of the $\bar{K}N, \pi\Sigma, \pi\Lambda, \eta\Sigma, \eta\Lambda, K\Xi$ channels. 
Conversely, we do not expect to find a signal for the $\Sigma(1385)(3/2^+)$, which according to Ref.~\cite{Capstick:1986ter} is a normal state of three quarks in its ground state, {with the spins aligned to give total spin 3/2}.

We shall compare our results with \xiu{the} experiment, which at present has very poor statistics~\cite{BESIII:2012jve}. 
For the moment, no signal of the $\Sigma(1385)$ is seen in the experiment, which dominates overwhelmingly in the $\pi\Lambda$ spectra of other reactions. There is some hint of the $\Sigma(1430)$ excitation, which we shall discuss; furthermore, our results give an idea of how much the statistics should be improved to see it clearly in the future.

Another related issue is the hypothetical $\Sigma(1380)(1/2^-)$, which has been claimed to be needed to improve agreement with experiments in some reactions (see an updated account of it in Ref.~\cite{Lyu:2026ack}), although in related reactions, like the $J/\psi \to \bar{\Lambda}\Sigma\pi$ reaction, one is able to explain deficiencies in the $\pi\Lambda$ mass distribution in the $\Sigma(1380)$ region by considering the $\pi\Sigma$ interaction without invoking the $\Sigma(1380)(1/2^-)$~\cite{Lyu:2026ums}.
Once again, the improved statistics in future runs of the experiment could help find out about that possible state. 
Based on our framework, if this resonance exists, it cannot be dynamically generated, because the interaction of coupled channels exhausts this structure with the $\Sigma(1430)(1/2^-)$ and the $\Sigma(1620)(1/2^-)$~\cite{Jido:2003cb}. 
And if it is not dynamically generated, it should not show up in this experiment. 

Consequently, this reaction serves as an excellent instrument to probe the nature of low-lying $\Sigma$ states. We hope the present work will stimulate future high-statistics studies of this process.

One theoretical work for this reaction is already available in Ref.~\cite{Huang:2024oai}, {proposing} a triangle mechanism, $J/\psi \to \bar{\Sigma}^* \Sigma$, followed by $\bar{\Sigma}^* \to \bar{\Lambda} \pi$ and \xiu{the} fusion of $\pi \Sigma$ to give $\pi \Lambda$. 
For some choices of $\bar{\Sigma}^*$, the triangle mechanism develops a triangle singularity, producing the $\Sigma(1430)$ in the $\pi^+ \Lambda$ mass distribution, but the $J/\psi \to \bar{\Sigma}^* \Sigma$ branching ratios are not known in this case {in} the RPP, which does not allow one to make quantitative predictions. 
The mechanism proposed there is very different to the one we propose here, but one thing in common is that the isospin breaking comes from \xiu{the noncancellation} of loops corresponding to intermediate particles belonging to the same isospin multiplets but different charges. 
Related to this issue, we find \xiu{it} appropriate to recall the work of Ref.~\cite{Jing:2019cbw} suggesting to measure the \xiu{isospin-violating} $J/\psi \to \phi \pi^0 \eta$ \xiu{reaction to observe} $a_0(980)\text{–}f_0(980)$ mixing. 
A peak was predicted in the $\phi \pi^0$ mass distribution at around $1385\text{ MeV}$, based on a triangle mechanism suggested in Ref.~\cite{Wu:2007jh}. The reaction was performed by the BESIII Collaboration in Ref.~\cite{BESIII:2023zwx}, where a peak was indeed seen around this energy. {However, the experimental team could attribute the strength of that peak to a ``non $\phi$" contribution.}
A theoretical work followed in Ref.~\cite{Li:2025rlj}, where the origin of the ``non $\phi$" contribution was identified, and the contribution of the triangle singularity was evaluated in absolute terms. Interestingly, the peak at the triangle singularity appeared at the same place as the experimental peak, {but its strength was a factor 40 smaller than the strength of the experimental peak.}

\section{Formalism}\label{sec2}
\begin{figure}[htp]
\centering
\begin{tabular}{cc}
  \includegraphics[width=0.24\textwidth]{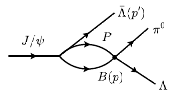} &
  \includegraphics[width=0.24\textwidth]{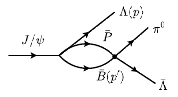} \\
  (a) & (b)  %
\end{tabular}
\caption{Rescattering diagrams for the decay $J/\psi \to \Lambda \bar{\Lambda} \pi^0$. (a) $PB$ interaction \xiu{and} (b) $\bar{P}\bar{B}$ interaction. \xiu{In parenthesis are} the momenta of the baryons. }
\label{fig:1}
\end{figure}
We follow the formalism developed in Ref.~\cite{Ikeno:2026zpv}. 
We start by using the fact that the $J/\psi$, being a $c\bar{c}$ state, is a scalar of SU(3) in the $u, d, s$ quarks. Then, a practical way to construct the most general SU(3) invariant structures with a baryon, an antibaryon\xiu{,} and one pseudoscalar meson is to write the SU(3) matrices,
\begin{equation}
P=
\begin{pmatrix}
\frac{1}{\sqrt{2}}\pi^0 + \frac{1}{\sqrt{3}} \eta & \pi^+ & K^+ \\[2mm]
\pi^- & -\frac{1}{\sqrt{2}} \pi^0 + \frac{1}{\sqrt{3}} \eta & K^0 \\[2mm]
K^- & \bar{K}^0 & -\frac{1}{\sqrt{3}} \eta
\end{pmatrix},
\label{eq:Pmatrix}
\end{equation}

\begin{equation}
B=
\begin{pmatrix}
\frac{1}{\sqrt{2}}\Sigma^0 + \frac{1}{\sqrt{6}} \Lambda & \Sigma^+ & p\\[2mm]
\Sigma^- & -\frac{1}{\sqrt{2}} \Sigma^0 + \frac{1}{\sqrt{6}} \Lambda & n \\[2mm]
\Xi^- & \Xi^0 & -\frac{2}{\sqrt{6}} \Lambda
\end{pmatrix},
\label{eq:Bmatrix}
\end{equation}

\begin{equation}
\bar{B}=
\begin{pmatrix}
\frac{1}{\sqrt{2}}\bar{\Sigma}^0 + \frac{1}{\sqrt{6}} \bar{\Lambda} & \bar{\Sigma}^+ & \bar{\Xi}^+ \\[2mm]
\bar{\Sigma}^- & -\frac{1}{\sqrt{2}} \bar{\Sigma}^0 + \frac{1}{\sqrt{6}} \bar{\Lambda }&  \bar{\Xi}^0\\[2mm]
\bar{p} & \bar{n} & -\frac{2}{\sqrt{6}} \bar{\Lambda}
\end{pmatrix}.
\label{eq:Bbarmatrix}
\end{equation}
Here, in $P$, we have considered the $\eta-\eta'$ mixing of Ref.~\cite{Bramon:1992kr}. We can write the SU(3) scalar structures in terms of the traces $\langle \bar{B} B P \rangle$, $\langle \bar{B} P B \rangle$, $\langle \bar{B} B \rangle \langle P \rangle$, $\langle \bar{B} P \rangle \langle B \rangle$, $\langle B P \rangle \langle \bar{B} \rangle$.
Since $B$ and $\bar{B}$ have zero trace, only $\langle \bar{B} B P \rangle$, $\langle \bar{B} P B \rangle$, and $\langle \bar{B} B \rangle \langle P \rangle$ structures survive. However, the $\langle \bar{B} B \rangle \langle P \rangle$ structure is suppressed for two reasons: one is because large $N_c$ counting penalizes extra traces~\cite{Manohar:1998xv}, and the other one {is} because if we consider only the octet of pseudoscalars in $P$, the trace of $P$ is also zero, hence it is only the small $\eta-\eta'$ mixing that makes its trace not zero. Thus, we are left with only two structures, $\langle \bar{B} B P \rangle$ and $\langle \bar{B} P B \rangle$, which are equivalent to those used in Refs.~\cite{He:2026mkf,Dai:2026zqn,Lyu:2026rsm}.

We assign weights $\tilde{A}$ and $\tilde{B}$ to these two structures, obtaining the Lagrangian~\cite{He:2026mkf}\xiu{,}
\begin{equation}
\mathcal{L} = \tilde{A} \langle \bar{B} \gamma^\mu \gamma_5 P B \rangle \Psi_\mu + \tilde{B} \langle \bar{B} \gamma^\mu \gamma_5 B P \rangle \Psi_\mu, 
\label{eq:Lagra}
\end{equation}
with $\Psi_\mu$ \xiu{being} the $J/\psi$ field.

We shall take into account the meson-baryon and meson-antibaryon interactions; hence we need all possible combinations having $\bar{\Lambda}$ meson-baryon and $\Lambda$ meson-antibaryon, which are given by
\begin{align}
\langle \bar{B} P B \rangle_{\bar{\Lambda}} 
&= \frac{\bar{\Lambda}}{\sqrt{6}} \Big( \pi^0 \Sigma^0 + \pi^+ \Sigma^- + \pi^- \Sigma^+ + K^+ \Xi^-  \notag \\
& + K^0 \Xi^0- 2 K^- p - 2 \bar{K}^0 n - \frac{2}{3\sqrt{2}} \eta \Lambda \Big), \label{eq:BPB_Lambar} 
\end{align}
\begin{align}
\langle \bar{B} B P \rangle_{\bar{\Lambda}} &= \frac{\bar{\Lambda}}{\sqrt{6}} \Big( \Sigma^0 \pi^0 + \Sigma^- \pi^+ + \Sigma^+ \pi^-- 2 \Xi^- K^+ \notag \\
&- 2 \Xi^0 K^0 + p K^- + n \bar{K}^0 - \frac{2}{3\sqrt{2}}\Lambda \eta \Big).
\label{eq:BBP_Lambar}
\end{align}
Similarly, isolating the terms with $\Lambda$, we have:
\begin{align}
\langle \bar{B} P B \rangle_{\Lambda} &= \frac{\Lambda}{\sqrt{6}} \Big( \bar{\Sigma}^0 \pi^0 + \bar{\Sigma}^+ \pi^- + \bar{\Sigma}^- \pi^+ + \bar{\Xi}^+ K^- \notag \\
&+\bar{\Xi}^0 \bar{K}^0- 2\bar{p} K^+ - 2\bar{n} K^0 - \frac{2}{3\sqrt{2}}\bar{\Lambda}\eta \Big), \label{eq:BPB_Lam}
\end{align}
\begin{align}
\langle \bar{B}BP \rangle_{\Lambda} &= \frac{\Lambda}{\sqrt{6}} \Big(
\pi^0 \bar{\Sigma}^0 + \pi^+ \bar{\Sigma}^- + \pi^- \bar{\Sigma}^+ - 2 K^- \bar{\Xi}^+ 
 \notag \\
&\qquad - 2 \bar{K}^0 \bar{\Xi}^0 + K^+ \bar{p}  + K^0 \bar{n} - \frac{2}{3\sqrt{2}}\, \eta \bar{\Lambda}
\Big). \label{eq:BBP_Lam}
\end{align}

The vertex structure $V \equiv -\mathcal{L}$ will be given by the matrix element:
\begin{align}
V \equiv -\bar{u}_\Lambda \gamma^\mu \gamma_5 v_{\bar{\Lambda}}(p') \epsilon_\mu \equiv \bar{u}_\Lambda \gamma^i \gamma_5 v_{\bar{\Lambda}}(p') \epsilon^i, 
\end{align}
where we have used \xiu{$\epsilon^0=0$} in the rest frame of $J/\psi$\xiu{,} and $u_\Lambda$, $v_{\bar{\Lambda}}$ are the spinors for the $\Lambda$ and $\bar{\Lambda}$.
Following Ref.~\cite{Ikeno:2026zpv} and keeping terms in the matrix element up to linear in momenta, we have:
\begin{align}
\bar{u}(p) \gamma^i \gamma_5 v(p')
&=
\chi^\dagger_r\left \{\frac{p^i}{2m_\Lambda} + \frac{p'^i}{2m_\Lambda} \notag \right. \\   
& \quad \quad \left. + i \epsilon_{ijk} \sigma_k \left( \frac{p'^j}{2m_\Lambda} - \frac{p^j}{2m_\Lambda} \right) \right \}\chi_{r'},
\label{eq:t_7}
\end{align}
with $\chi_r$, $\chi_{r'}$ bispinors for $\Lambda$ and $\bar{\Lambda}$ respectively.

As mentioned above, we shall have {the types of diagrams} that we depict in Fig.~\ref{fig:1}.
The explicit separation of terms in Eq.~\eqref{eq:t_7} is appropriate for our problem. Indeed, since the $PB \to \pi^0 \Lambda$ and $\bar{P}\bar{B} \to \pi^0 \bar{\Lambda}$ amplitudes are in $S$-wave, in the diagrams of Type (a) in Fig.~\ref{fig:1} the terms with $p^i$ in Eq.~\eqref{eq:t_7}  will not contribute, and, similarly, in the diagrams of Type (b) the terms with $p'^i$ will not contribute.

\begin{figure}[htbp]
\begin{center}
	\begin{tabular}{cc}
		\includegraphics[width=0.24\textwidth]{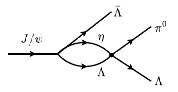}&
		\includegraphics[width=0.24\textwidth]{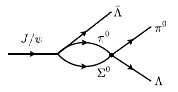}
	\end{tabular}
	\begin{tabular}{cc}
		\includegraphics[width=0.24\textwidth]{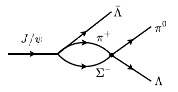}&
		\includegraphics[width=0.24\textwidth]{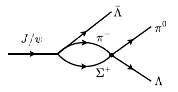}
	\end{tabular}
	\begin{tabular}{cc}
		\includegraphics[width=0.24\textwidth]{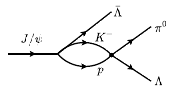}&
		\includegraphics[width=0.24\textwidth]{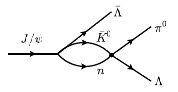}
	\end{tabular}
	\begin{tabular}{cc}
		\includegraphics[width=0.24\textwidth]{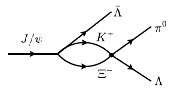}&
		\includegraphics[width=0.24\textwidth]{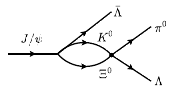}
	\end{tabular}
  \caption{Diagrams contributing to $J/\psi \to \Lambda \bar{\Lambda} \pi^0$ where $\bar{\Lambda}$ is a spectator and the $MB$ pairs interact.}
  \label{fig:2}
\end{center}
\end{figure}

\begin{figure}[htbp]
\begin{center}
	\begin{tabular}{cc}
		\includegraphics[width=0.24\textwidth]{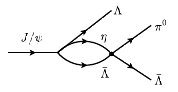}&
		\includegraphics[width=0.24\textwidth]{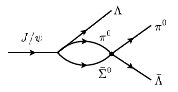}
	\end{tabular}
	\begin{tabular}{cc}
		\includegraphics[width=0.24\textwidth]{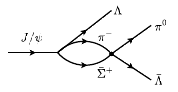}&
		\includegraphics[width=0.24\textwidth]{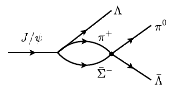}
	\end{tabular}
	\begin{tabular}{cc}
		\includegraphics[width=0.24\textwidth]{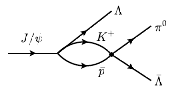}&
		\includegraphics[width=0.24\textwidth]{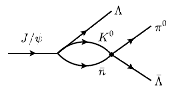}
	\end{tabular}
	\begin{tabular}{cc}
		\includegraphics[width=0.24\textwidth]{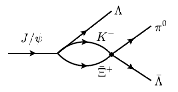}&
		\includegraphics[width=0.24\textwidth]{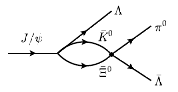}
	\end{tabular}
  \caption{Diagrams contributing to $J/\psi \to \Lambda \bar{\Lambda} \pi^0$ where $\Lambda$ is a spectator and the $\bar{M}\bar{B}$ pairs interact.}
  \label{fig:3}
\end{center}
\end{figure}

The diagrams contributing to our reaction are depicted in Figs.~\ref{fig:2} and \ref{fig:3}. 
In Fig.~\ref{fig:2}, we gather all diagrams where $\bar{\Lambda}$ is a spectator and the meson-baryon pairs interact, while in Fig.~\ref{fig:3}, we keep the terms where $\Lambda$ is a spectator and the $\bar{M}\bar{B}$ pairs interact.

Note that, according to Eqs.~\eqref{eq:BPB_Lambar}--\eqref{eq:BBP_Lam}, the diagrams of Figs.~\ref{fig:2} and \ref{fig:3} have different weights for the $\langle \bar{B} P B\rangle$ and $\langle \bar{B} B P \rangle$ structures. Accordingly, we get the final structure for the decay amplitudes, $t_1$ for the diagrams of Type (a) and $t_2$ for the diagrams of Type (b), given by
\begin{align}
t_{1} & = -\frac{1}{3 \sqrt{3}}(\widetilde{A}+\widetilde{B}) G_{\eta \Lambda} t_{\eta \Lambda, \pi^{0} \Lambda} \notag \\
& +\frac{(\widetilde{A}+\widetilde{B})}{\sqrt{6}}\{G_{\pi^{0} \Sigma^{0}} t_{\pi^{0} \Sigma^{0}, \pi^{0} \Lambda}+G_{\pi^{+} \Sigma^{-}} t_{\pi^{+} \Sigma^{-}, \pi^{0} \Lambda}\notag \\
&\qquad\qquad\quad +G_{\pi^{-} \Sigma^{+}} t_{\pi^{-} \Sigma^{+}, \pi^{0} \Lambda}\}\notag \\
& +\frac{(-2 \widetilde{A}+\widetilde{B})}{\sqrt{6}}\left\{G_{K^{-} p} t_{K^{-} p, \pi^{0} \Lambda}+G_{\bar{K}^{0} n} t_{\bar{K}^{0} n, \pi^{0} \Lambda}\right\}\notag \\
& +\frac{(\widetilde{A}-2 \widetilde{B})}{\sqrt{6}}\left\{G_{K^{+} \Xi^{-}} t_{K^{+} \Xi^{-}, \pi^{0} \Lambda}+G_{K^{0} \Xi^{0}} t_{K^{0} \Xi^{0}, \pi^{0} \Lambda}\right\}, 
\label{eq:t_8}
\end{align}
\begin{align}
t_{2} & =  -\frac{1}{3 \sqrt{3}}(\widetilde{A}+\widetilde{B})G_{\eta \bar{\Lambda}} t_{\eta \Lambda, \pi^{0} \Lambda}\notag \\
& +\frac{(\widetilde{A}+\widetilde{B})}{\sqrt{6}}\{G_{\pi^{0} \bar{\Sigma}^{0}} t_{\pi^{0} \Sigma^{0}, \pi^{0} \Lambda}+G_{\pi^{-} \bar{\Sigma}^{+}} t_{\pi^{+} \Sigma^{-}, \pi^{0} \Lambda}\notag \\
&\qquad\qquad\quad +G_{\pi^{+} \bar{\Sigma}^{-}} t_{\pi^{-} \Sigma^{+}, \pi^{0} \Lambda}\} \notag \\
& +\frac{(-2 \widetilde{A}+\widetilde{B})}{\sqrt{6}}\left\{G_{K^{+} \bar{p}} t_{K^{-} p, \pi^{0} \Lambda}+G_{K^{0} \bar{n}} t_{\bar{K}^{0} n, \pi^{0} \Lambda}\right\}\notag \\
& +\frac{(\widetilde{A}-2 \widetilde{B})}{\sqrt{6}}\left\{G_{K^{-} \bar{\Xi}^{+}} t_{K^{+} \Xi^{-}, \pi^{0} \Lambda}+G_{\bar{K}^{0} \bar{\Xi}^{0}} t_{K^{0} \Xi^{0}, \pi^{0} \Lambda}\right\}.
\label{eq:t_9}
\end{align}
Note that in Eq.~\eqref{eq:t_9} we have already taken the freedom to convert $t_{\bar{M}\bar{B}, \bar{M}\bar{B}}$ into the corresponding $t_{MB, MB}$, which are {the} amplitudes evaluated in the literature. \xiu{However, the argument of the function in $t_1$ is $M_{\rm inv}(\pi^0\Lambda)$, while for $t_2$ it is $M_{\rm inv}(\pi^0\bar{\Lambda})$.} In Eqs.~\eqref{eq:t_8} and \eqref{eq:t_9}, the amplitudes and meson-baryon loop functions are taken directly from Refs.~\cite{Oset:1997it,Oset:2001cn}.

Eqs.~\eqref{eq:t_8} and \eqref{eq:t_9} still do not incorporate the spinor structure. We incorporate it into the final amplitudes $\tilde{t}_1, \tilde{t}_2$ given by:
\begin{align}
\tilde{t}_1 = t_1 \langle \chi_r | \frac{p'_i}{2m_1} + i \epsilon_{ijk} \sigma_k \frac{p'_j}{2m_1} | \chi_{r'} \rangle \epsilon^i, 
\label{eq:t_10}
\end{align}
\begin{align}
\tilde{t}_2 = t_2 \langle \chi_r | \frac{p_i}{2m_3} - i \epsilon_{ijk} \sigma_k \frac{p_j}{2m_3} | \chi_{r'} \rangle \epsilon^i, 
\label{eq:t_11}
\end{align}
{with the label assignment $\bar{\Lambda}(1), \pi^0(2), \Lambda(3)$.} And summing $\tilde{t}_1$ and $\tilde{t}_2$ we obtain:
\begin{align}
t^{\text{tot}} &= \tilde{t}_1 + \tilde{t}_2 \notag  \\
&= \frac{1}{2M_\Lambda} \langle \chi_r | (t_1 p'_i + t_2 p_i) + i \epsilon_{ijk} \sigma_k (t_1 p'_j - t_2 p_j) | \chi_{r'} \rangle \epsilon^i.\label{eq:t_12}
\end{align}
Finally, when we sum $|t^{\text{tot}}|^2$ over final baryon polarizations and average over the initial $J/\psi$ polarization, we obtain
\begin{align}
\overline{\sum}\sum |t^{\text{tot}}|^2 
&= \frac{1}{2 M_\Lambda^2} \Big[ |t_1|^2 |\mathbf{p}^\prime|^2 + |t_2|^2 |\mathbf{p}|^2 \notag  \\
&\qquad\qquad- \frac{2}{3}\mathrm{Re}(t_1 t_2^*) (\mathbf{p} \cdot \mathbf{p}^\prime) \Big].
\label{eq:t_13}
\end{align}

The mass distribution is given by the master formula of the RPP~\cite{ParticleDataGroup:2024cfk}, adapted to the Mandl-Shaw normalization of the fields~\cite{Mandl:2010qft}:
\begin{equation}
\frac{d^2 \Gamma}{dM_{12}dM_{23}} = \frac{1}{(2\pi)^3} \frac{(2 {M_{\Lambda}})^2}{32 m_\psi^3} \overline{\sum}\sum |t^{\text{tot}}|^2 2M_{12} 2M_{23}.
\label{eq:t_14}
\end{equation}
In Eq.~\eqref{eq:t_13}, we see that we need $p$, $p^\prime$ and $\vec{p} \cdot \vec{p}^{~\prime}$, and $t_1$ depends on $M_{23} = M_{\pi^0\Lambda}$ and $t_2$ on $M_{12} = M_{\pi^0\bar{\Lambda}}$. 
By integrating over $M_{23}$ with the limits of the RPP, we obtain $d\Gamma/dM_{12} = d\Gamma/dM_{\pi^0\bar{\Lambda}}$, which is the same as $d\Gamma/dM_{23} = d\Gamma/dM_{\Lambda\pi^0}$.
We have:
\begin{equation}
|p| =
\frac{\lambda^{1/2}(m_{J/\psi}^2, {M_\Lambda^2}, M_{12}^2)}{2m_{J/\psi}},
\quad
|p^\prime| =
\frac{\lambda^{1/2}(m_{J/\psi}^2, {M^2_{\bar{\Lambda}}}, M_{23}^2)}{2m_{J/\psi}}.\label{eq:t_1516}
\end{equation}
Analogously,
\begin{equation}
\vec{p} \cdot \vec{p}^{~\prime} = \vec{p}_1 \cdot \vec{p}_3 = \frac{1}{2} (2m_1^2 + 2E_1 E_3 - M_{13}^2), \label{eq:t_17}
\end{equation}
where
\begin{equation}
 E_1 = \frac{m_{J/\psi}^2 + {M_{\bar{\Lambda}}^2} - M_{23}^2 }{2 m_{J/\psi}},
\quad
 E_3 = \frac{m_{J/\psi}^2 + {M_{\Lambda}}^2 - M_{12}^2 }{2 m_{J/\psi}}.
\end{equation}
Also, we need $M_{13}$ in Eq.~\eqref{eq:t_17}, which can be written in terms of $M_{12}$, $M_{23}$ since:
\begin{equation}
M_{12}^2 + M_{13}^2 + M_{23}^2 = m_{J/\psi}^2 + m_1^2 + m_2^2 + m_3^2.
\end{equation}

\section{ Results }\label{sec3}
\begin{figure}
    \centering
    \includegraphics[width=1\linewidth]{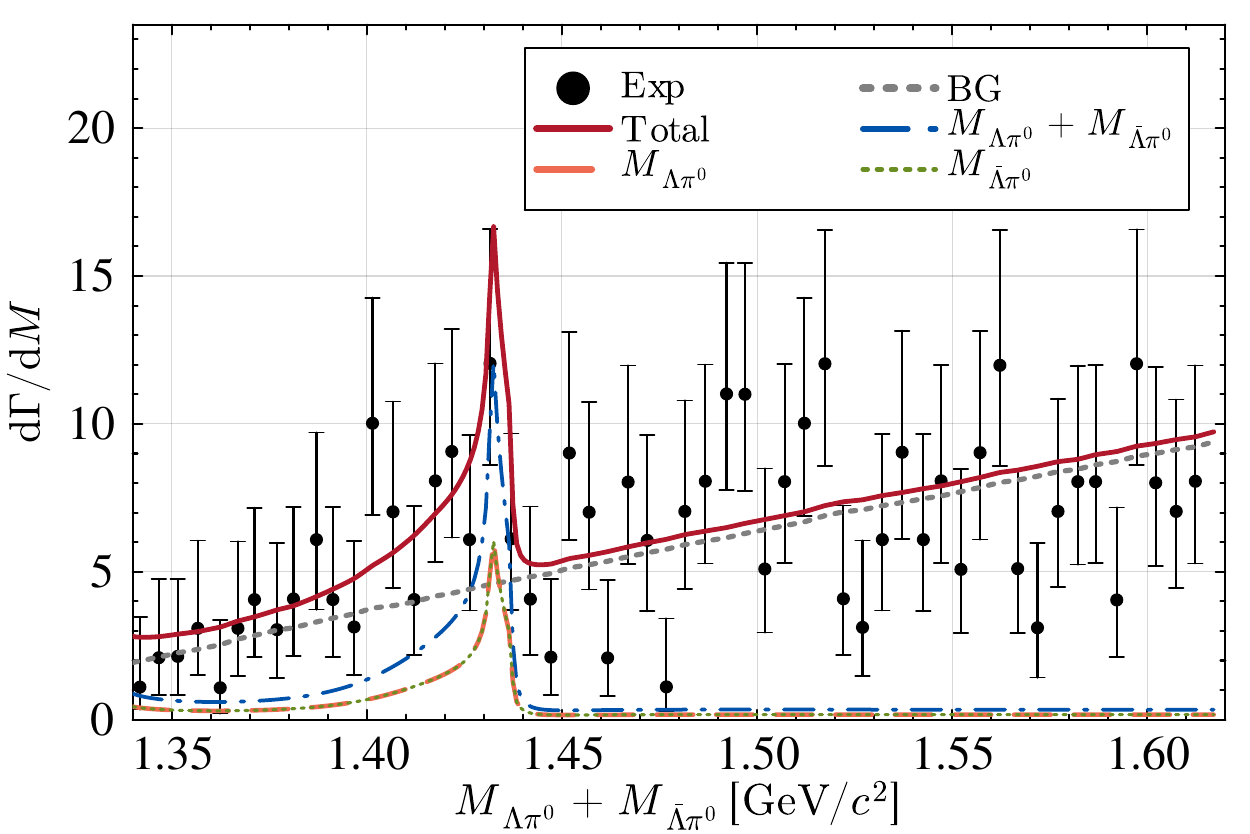}
    \caption{Invariant mass distribution of $\Lambda\pi^0$ and $\bar{\Lambda}\pi^0$ for the $J/\psi \to \Lambda \bar{\Lambda} \pi^0$ reaction. The solid red line represents the total fit to the data, including the adjusted smooth background (dashed line). The data are taken from Ref.~\cite{BESIII:2012jve}.}
    \label{fig:4}
\end{figure}
The results are shown in Fig.~\ref{fig:4} compared with the BESIII data~\cite{BESIII:2012jve}. 
We have added a smooth background, also taken from Ref.~\cite{BESIII:2012jve}, to better reproduce the data.
We have two parameters $\tilde{A}$ and $\tilde{B}$, and we take the ratio 
$\tilde{A}/\tilde{B} = 1.44$ from \xiu{Ref.}~\cite{He:2026mkf}. 
Thus, we only have a global normalization factor that we adjust to the data.
As discussed above, only the dynamically generated resonances, stemming from the meson-baryon interaction, are expected to appear. 
In our case, this is the $\Sigma(1430)$ state. It is not easy to claim that the data show a signal for this resonance. 
However, there is one data point around $1430 \text{ MeV}$ standing above the rest, and a subsequent drop in the data, which our calculation reproduces very well.
It is just an indication, but \xiu{it is} sufficiently suggestive to encourage precise measurements, which can be done in the future with the ongoing accumulation of $J/\psi$ decays in BESIII.

\begin{figure}
    \centering
    \includegraphics[width=1\linewidth]{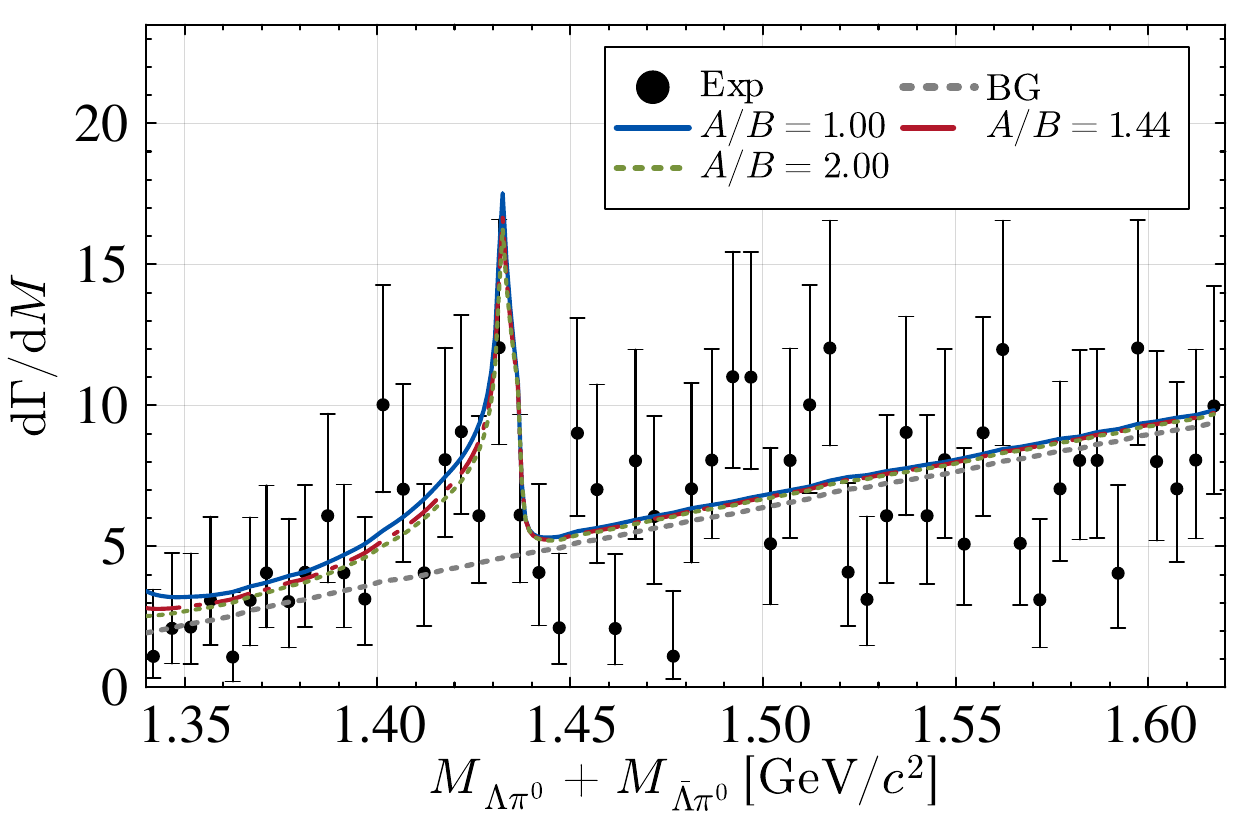}
    \caption{Invariant mass distribution of $\Lambda\pi^0$ and $\bar{\Lambda}\pi^0$ with different ratios of $\tilde{A}/\tilde{B}$.}
    \label{fig:5}
\end{figure}

In order to quantify uncertainties of our results, we show in Fig.~\ref{fig:5} the mass distributions that we obtain with reasonable changes of the $\tilde{A}/\tilde{B}$ ratio, keeping the value of $\tilde{A}$ as in Fig.~\ref{fig:4}.
We can see that setting $\tilde{A}/\tilde{B}=1$ or $\tilde{A}/\tilde{B}=2$, which differ appreciably from the value $1.44$ assumed in Fig.~\ref{fig:4}, the results of the mass distributions change very little, giving support to the stability of these results.

It is interesting to observe that, contrary to what appears in other experiments where the $\Sigma(1385)$ shows up as the largely dominant contribution {(see for instance the $J/\psi \to \Lambda \pi \bar{\Sigma}$ reaction~\cite{Lyu:2026ums, BESIII:2023syz})}, even with the limited statistics {in Fig.~\ref{fig:4}}, one can claim that there is no signal of the $\Sigma(1385)$, as was already noted in the experimental paper~\cite{BESIII:2012jve}. 

In summary, these observations demonstrate that isospin violating reactions are an excellent tool to differentiate between genuine $qqq$ states and dynamically generated resonances. 
The present case seems to be one example of this rule, with clear clues that the $\Sigma(1430)$ state is seen, while the $\Sigma(1385)$ does not show up.

\section{ Conclusions }\label{sec4}
We have studied the isospin-violating $J/\psi \to \bar{\Lambda} \Lambda \pi^0$ reaction using a theoretical framework that has been successfully applied to similar reactions. 
The approach starts by considering all possible terms of baryon-antibaryon-pseudoscalar meson that form a scalar of $\text{SU}(3)$ in the $u, d, s$ quarks. From them, we isolate the terms that have a $\bar{\Lambda}$\xiu{,} allowing the remaining meson-baryon pairs to interact and ultimately produce the $\pi^0\Lambda$ state. 
Similarly, we also consider the terms that have a $\Lambda$ and let the meson-antibaryon interact to give finally the $\pi^0\bar{\Lambda}$.
The approach leads to a $\bar{\Lambda} \Lambda \pi^0$ final state thanks to the interactions which mildly violate isospin due to the different masses of particles in the same isospin multiplets. 
In our approach, only the final-state interactions lead to the desired final states, and this is a filter for the observation of dynamically generated resonances which come from the meson-baryon interaction.
With the limited statistics that the BESIII experiment provides, one can see features in the data that are well reproduced by the $\Sigma(1430)(1/2^-)$ state, a dynamically generated state within the chiral unitary approach, while the data do not show any signal of the $\Sigma(1385)(3/2^+)$ state, a conventional ground state of three quarks with spin $3/2$. 
The findings of the present study should stimulate further investigation into this reaction with improved statistics, which should be possible in the near future at BESIII.

\section*{Acknowledgments}
This work was supported by the National Key R\&D Program of China (Grant No. 2024YFE0105200), the Natural Science Foundation of Henan (Grant No. 252300423951), and the Zhengzhou University Young Student Basic Research Projects for PhD students (Grant No. ZDBJ202522). Yu-Shan Ren and Wen-Tao Lyu also acknowledge the support from the China Scholarship Council.
This work is also partly supported by the Spanish Ministerio de Economia y Competitividad~(MINECO) and European FEDER funds under Contracts No. FIS2017-84038-C2-1-PB \xiu{and No. PID2020--112777 GB-I00} and by Generalitat Valenciana under \xiu{Contract No.} PROMETEO/2020/023. This project has received funding from the European Union Horizon 2020 research and innovation program under the program H2020-INFRAIA-2018-1, \xiu{Grant Agreement} No. 824093 of the STRONG-2020 project.
\xiu{Research was} partially supported by \xiu{Grant No.} PID2023-147458NB-C21 funded by MCIN/AEI/ 10.13039/501100011033 and by the European Union.

\bibliography{ref.bib}
\end{document}